\documentstyle[aps,preprint]{revtex}
\def\lsim{\mathrel{\raise2pt\hbox to 8pt{\raise -5pt\hbox{$\sim$}\hss{$<$}}}}
\title{Pseudo-spin as a
Relativistic Symmetry}
\author{Joseph N.\ Ginocchio}
\address{{\it T-5, MS B283, Theoretical Division, Los Alamos National
Laboratory\\ Los Alamos, NM 87545\\ }}
\begin{document}
\draft
\maketitle
\begin{abstract}
\noindent We show that pseudo - spin symmetry in nuclei could arise from
nucleons moving in a relativistic mean
field which has an attractive scalar and repulsive vector potential nearly
equal in magnitude.\\

\end{abstract}
\pagebreak

Almost thirty years ago a quasi-degeneracy was observed in heavy nuclei
between single-nucleon doublets with quantum numbers
($n_r$, $\ell$,\ j= $\ell$ + 1/2) and ($n_r$-1, $\ell+2$, j= $\ell$ + 3/2)
where $n_r$,$\ell$, and
j are the single nucleon radial, orbital, and total angular momentum
quantum numbers,
respectively \cite{kth,aa}. These authors defined a "pseudo" orbital
angular momentum
$\tilde{\ell}$ = $\ell$ + 1;
for example, $(n_r s_{1/2},(n_r-1) d_{3/2})$ will have $\tilde{\ell}= 1$ ,
$(n_r p_{3/2},(n_r-1)
f_{5/2})$ will have $\tilde{\ell}= 2$, etc. Then these doublets are almost
degenerate with
respect to "pseudo" spin, $\tilde s$ = 1/2, since j = $\tilde{\ell}\ \pm
\tilde s$ for the two states in the doublet.
This symmetry has been used to explain
a number of phenomena in nuclear structure \cite{bohr} including most
recently the identical rotational bands observed in nuclei
\cite{ben}. Despite this long history of pseudo-spin
symmetry \cite{mosh,jpd1}, the origin of this symmetry had eluded
explanation. Recenty, it was shown
\cite{jpd} that relativistic mean field theories predict the correct
spin-orbit splitting \cite{wal}. In this paper we identify
a possible reason for this; namely, that the symmetry arises from the near
equality in magnitude of an attractive scalar, -V$_s$, and
repulsive vector, V$_v$, relativistic mean fields, V$_s \sim\ $V$_v$, in
which the nucleons move. Such a near equality of mean fields
follows from relativistic field theories with interacting nucleons and
mesons \cite {wal}, with nucleons interacting via Skyrme-type
interactions \cite{mad},
and from QCD sum rules \cite{furn}.

A nucleon moving in a spherical field has the total angular momentum j, its
projection on the z-axis, m, and ${\hat {\kappa}} =
-{\hat {\beta}}( {\hat {\sigma}}\cdot{\hat {L}} + 1) $ conserved, where
${\hat {\beta}}$ is the Dirac matrix \cite {mul}.
The eigenvalues of
$\hat {\kappa}$ are
$\kappa =
\pm$(j + 1/2); - for aligned spin ($s_{1/2},p_{3/2},$ etc.) and + for
unaligned spin ($p_{1/2},
d_{3/2}, $ etc.). Hence, we use the quantum number $\kappa$ since it is
sufficient to label the
orbitals. The Dirac equation for the single - nucleon radial wavefunction
($g_{\kappa}, f_{\kappa}$) in
dimensionless units is given by \cite{mul}
\begin{equation}
\ [{d\over{dr}}+ {{ 1 + \kappa}\over r}\ ] g_{\kappa} = \ [ 2 - E - V(r) \
] f_{\kappa}
\label {eq_d1}
\end{equation}
\begin{equation}
\ [{d\over{dr}}+ {{ 1 - \kappa}\over r}\ ] f_{\kappa} = \ [E - \Delta(r) \
]g_{\kappa},
\label {eq_d2}
\end{equation}
where r is the radial coordinate in units of length ${\hbar c}/{mc^2}$,
$V(r)$ = (V$_v(r)$ + V$_s(r)$) / $mc^2$,
$\Delta(r)$ = (V$_s(r)$ - V$_v(r)$) / $mc^2$, and $E$ is the binding energy
($E > 0$) of the nucleon in units of the nucleon mass,
$mc^2$. First we show that, in the limit of equality of the magnitude of
the vector and scalar potential, $\Delta( r )$ = 0, pseudo
spin is exactly conserved. To do this, we solve for $g_{\kappa}$ in (\ref
{eq_d2}) and substitute into (\ref{eq_d1}), obtaining the
second order differential equation for $f_{\kappa}$,
\begin{equation}
\ [{{d^2}\over{dx^2}}+ {2\over x}{{d}\over{dx}}-{{ \tilde {\ell} ( \tilde
{\ell} + 1)}\over {x^2}}\ + ( V(r) - 2 + E )]
f_{\kappa} = 0,
\label {eq_s}
\end{equation}
where $x = \sqrt E\ r$ and

%\pagebreak
\begin{equation}
\tilde {\ell} = \kappa - 1, \kappa > 0;
\tilde {\ell} = - \kappa, \kappa < 0,
\label {eq_k}
\end{equation}
which agrees with the original definition of the pseudo-orbital angular
momentum \cite {kth,aa}. For example,
for $(n_r s_{1/2},(n_r-1) d_{3/2})$, $\kappa =
-1$ and $2$, respectively giving $\tilde {\ell} =1$ in both cases.
Furthermore, the physical
significance of $\tilde {\ell} $ is revealed; it is the "orbital angular
momentum" of the lower
component of the Dirac wavefunction.

Equation (\ref {eq_s}) is a
Schr\"{o}dinger equation with an attractive potential $V$ and binding
energy $2 - E$ which depends only on the pseudo-orbital
angular momentum,
$\tilde {\ell}$, through the pseudo - rotational kinetic energy, ${{ \tilde
{\ell} ( \tilde
{\ell} + 1)}\over {x^2}}$, and not on
$\kappa$. Hence the eigenenergies and eigenfunction component $f_{\kappa}$
do not depend on
$\kappa$ but only on $ \tilde {\ell}$. Thus the doublets with the same $
\tilde {\ell}$ but different
$\kappa \ (\kappa = \tilde {\ell} + 1$ and $\kappa = - \tilde {\ell})$ will
be degenerate, producing pseudo - spin symmetry.

However, in this limit, there will not be any bound Dirac valence states,
only Dirac sea states, which contradicts reality.
Is it possible that we can have bound valence states $\it {and}$
quasi-degeneracy for a small $\Delta( r )$? To answer that question
we look at two examples. First is the spherical Coulomb potential and the
second is the spherical potential well.

The spherical Coulomb potential for arbitrary scalar and vector fields,
$V_{s,v}(r) = \alpha_{s,v}/ r$, can be solved
analytically \cite {mul}. The valence eigenenergies are given by,
\begin{equation}
E_{n,\kappa} = {{4(n + \lambda)^2 \ \big [ 1 - {\sqrt{(1 - (\alpha \delta)/
(n + \lambda)^2}}\
\big ] - 2 \delta (\alpha -\delta)}
\over{(\alpha -
\delta)^2 + 4(n + \lambda)^2 } }, \label {eq_e}
\end{equation}
where $n$ is the principal quantum number, n = 1, 2,\dots, $\alpha =
\alpha_s + \alpha_v$, $\delta = \alpha_s - \alpha_v$,
and $\lambda = |\kappa|(\sqrt{1 + {{\alpha \delta}\over {{\kappa}^2}}} - 1
)$. The allowed values of
$\kappa $ are $\kappa = \pm1,\pm2,\dots,\pm(n-1), -n. $ The dependence on
$\kappa$ is in $\lambda$. If the scalar and vector
potential are equal, $\delta = 0$, then the binding energy vanishes,
$E_{n,\kappa} = 0$, and hence no bound valence states as stated
earlier. For $\delta$ small,

\begin{equation}
E_{n,\kappa} \approx {{{\delta}^2}\over {2n^2}}( 1 + {{\alpha \delta}\
(2n^2(|\kappa| - 2) - {\alpha}^2) \over{|\kappa| n^2(4n^2 + {\alpha}^2)} }\
+ \dots \ ), \label {eq_app}
\end{equation}
Thus the pseudo-spin symmetry is broken in third order in $\delta$. We
notice that the breaking decreases as $n$
increases and, for a given $n$, the state with the largest $|\kappa|$
(which means the pseudo - spin partner with $\kappa > 0$) will
have the largest binding energy. Thus pseudo - spin quasi-degeneracy
coexists with an infinite
number of bound valence Dirac states.

However, the Coulomb potential is not realistic for nuclei and,
furthermore, the Coulomb potential has higher degeneracies than
pseudo-spin since the energies depend only on n, and not $\kappa$ in the
lowest order. For these reasons we turn now to the spherical
potential well:

\begin{equation}
V_{s,v}(r) = V_{s,v} > 0, r < R;
V_{s,v}(r) = 0, r > R;
\label {eq_w}
\end{equation}
The soution of the Dirac equation (\ref {eq_d1},\ref {eq_d2}) is given in
terms of spherical Bessel functions for $r < R$,

\begin{equation}
g_{\kappa} = A\ {\bf j}_{{\tilde {\ell}}+ {\tilde {\epsilon}}}(z), \
f_{\kappa} = {{\tilde {\epsilon} A k}\over{2 - E - V}}\
{\bf j}_{{\tilde {\ell}}}(z)\ , r < R ,
\label {eq_bi}
\end{equation}
and modified spherical
Bessel functions for $r > R$ \cite {mul}, \begin{equation}
g_{\kappa} = \bar A\ {\bf k}_{{\tilde {\ell}}+ \tilde {\epsilon}}(y), \
f_{\kappa} = -{{\bar A K}\over{2 - E }}\
{\bf k}_{{\tilde {\ell}}}(y)\ , r > R,
\label {eq_bo}
\end{equation}
where $z = kr,\ y = Kr,\ $ and the wave numbers are given by
\begin{equation}
k^2 = (\Delta - E)( 2 - E - V) > 0, \ K^2 = E (2 - E) > 0, \label {eq_k}
\end{equation}
where $\tilde {\epsilon} = \kappa / |\kappa|$ is the pseudo - helicity,
since $j = \tilde {\ell} +
\tilde {\epsilon}
\ 1/2$, and the eigenvalues of $\tilde {\epsilon}$ are $\pm 1$.

The two solutions must match at the boundary, $r = R$, leading to the two
conditions which determine the eigenvalues for the same
$\tilde {\ell}$, but different $ {\kappa}$: \begin{equation}
{{Z_{\kappa}{\bf j}_{{\tilde
{\ell}} + 1}(Z_{\kappa})}\over {{\bf j}_{\tilde{\ell}}}
(Z_{\kappa})} = - {{Y_{\kappa}(\Delta - E_{\kappa})\ {\bf k}_{{\tilde
{\ell}} + 1}(Y_{\kappa})\over {E_{\kappa}
\ {\bf k}_{\tilde{\ell}}} (Y_{\kappa})}}, \kappa > 0 ,
\label {eq_e}
\end{equation}
\begin{equation}
{{Z_{\kappa}{\bf j}_{{\tilde
{\ell}} - 1}(Z_{\kappa})}\over {{\bf j}_{\tilde{\ell}}}
(Z_{\kappa})} = {{Y_{\kappa}(\Delta - E_{\kappa})\ {\bf k}_{{\tilde {\ell}}
- 1}(Y_{\kappa})\over {E_{\kappa}\
{\bf k}_{\tilde{\ell}}} (Y_{\kappa})}}, \kappa < 0,
\label {bessel}
\end{equation}
where $Z = kR$ and $Y = KR$. Since there are two different equations for the
states with the same $\tilde{\ell}$ but different $\tilde {\kappa}$, the
eigenenergies of these two
different states will be different in general.

Using the recurrence relations between Bessel functions \begin{equation}
{{Z{\bf j}_{{\tilde{\ell}} + 1}(Z)} = (2\tilde{\ell} + 1)\ {\bf
j}_{{\tilde{\ell}}}(Z) - Z{\bf j}_{{\tilde{\ell} - 1}}(Z);
\ Y{\bf k}_{{\tilde{\ell}} + 1}(Y)} = (2\tilde{\ell} + 1)\ {\bf
k}_{\tilde{\ell}}(Y) + Y{\bf k}_{{\tilde{\ell}} - 1}(Y),
\end{equation}
we can eliminate ${\bf j}_{{\tilde {\ell}} - 1},{\bf k}_{{\tilde {\ell}} -
1}$ and rewrite these
equations as,
\begin{equation}
-{{Z_{\kappa}{\bf j}_{{\tilde
{\ell}} + 1}(Z_{\kappa})}\over {{\bf j}_{\tilde{\ell}}}
(Z_{\kappa})} = {{Y_{\kappa}(\Delta - E_{\kappa})\ {\bf k}_{{\tilde {\ell}}
+ 1}(Y_{\kappa})\over {E_{\kappa}\
{\bf k}_{\tilde{\ell}}} (Y_{\kappa})}}, \ \kappa > 0 , \label {bessel>}
\end{equation}
\begin{equation}
-{{Z_{\kappa}{\bf j}_{{\tilde
{\ell}} + 1}(Z_{\kappa})}\over {{\bf j}_{\tilde{\ell}}}
(Z_{\kappa})} = {{Y_{\kappa}(\Delta - E_{\kappa})\ {\bf k}_{{\tilde {\ell}}
+ 1}(Y_{\kappa})\over {E_{\kappa}\
{\bf k}_{\tilde{\ell}}} (Y_{\kappa})}}- (2\tilde{\ell} + 1) {\Delta \over
E_{\kappa}}, \ \kappa < 0,
\label {bessel<}
\end{equation}
thereby displaying the fact that the equations become identical for
$\Delta = 0$ producing the
pseudo-spin degeneracy but, as we shall see, no Dirac valence bound states.

In Fig.1 we plot the left - hand side (LHS) of
(\ref{bessel>},\ref{bessel<}) as a
function of $Z$. The LHS decreases from a value of zero at $Z$ = 0 to
negative infinity at $Z^{(0)}_{1,\tilde {\ell}}$ ,
where $Z^{(0)}_{n,\tilde {\ell}}$
is the
$n$th zero of the spherical Bessel function, $j_{\tilde {\ell} }(
Z^{(0)}_{n, \tilde
{\ell}}) = 0$, with $Z = 0$ corresponding to $n = 0$. The LHS then becomes
discontinuous at this
point, and for $ Z > Z^{(0)}_{1,\tilde {\ell}} $ it decreases from positive
infinity to
zero at
$Z^{(0)}_{1,\tilde {\ell} + 1}$, and then negative infinity at
$Z^{(0)}_{2,\tilde
{\ell}}$ and so on. We call the region with $ Z^{(0)}_{n,\tilde {\ell}} \le
Z < Z^{(0)}_{n+1,\tilde
{\ell}}$ the $nth$ branch.

On the other hand, the right-hand side (RHS) of both (\ref{bessel>}) and
(\ref{bessel<}) increases
monotonically, as illustrated in Fig. 2. The eigenvalues are determined by
the points of intersections in the nth branch,
$Z_{n,\tilde
{\kappa}}$, giving the valence eigenergies,
\pagebreak
$$
E_{n_r,{\kappa}} =\frac { 2 - V + \Delta}{2} - \sqrt{ \big (\frac { 2 - V -
\Delta}{2}\big )^2 +
\big (\frac {Z_{n,{\kappa}}}{R} \big )^2},
$$
\begin{equation}
n_r = n - 1, \kappa > 0, n_r = n, \kappa < 0, \label {energy}
\end{equation}
where the radial quantum number will become clear subsequently. The RHS of
(\ref{bessel>}) increases
from zero at $Z = 0\ (E =\Delta)$ to infinity at
$Z = Z_{max} = \sqrt{\Delta ( 2 - V)}\ R$ $(E=0)$ ( see (\ref {energy}) ).
However the RHS of
(\ref{bessel<}) is smaller by an amount $(2\tilde {\ell} + 1){\Delta \over
E}$, and increases from
$-(2\ell + 1)$ at $Z = 0 \ (E = \Delta) $ to a constant at $Z = Z_{max}$
$(E=0)$.
This means that for
$ {\kappa} < 0$, there will be a bound state for each branch in Fig. 2 as
long as
$Z_{n,{\kappa}} < Z_{max}$. Furthermore the upper component for
$ {\kappa} < 0$, ${\bf j}_{{\tilde {\ell}} - 1}$, has a zero in each of
these branches and thus
the radial quantum number is then
$n_r = n = 0,1,\dots$. However, for $ {\kappa > 0} $, there will not be a
bound state for $n = 0$
since $\frac{\Delta}{E} = 1 $ at $Z = 0$, and thus the two curves intersect
only at $Z = 0$, which means $k = 0$ and hence
there is no bound state for $n= 0$ ( see
(\ref{eq_bi})). However, there will be a bound state for all the other
branches in Fig. 2 as long as
$Z_{n,{\kappa}} < Z_{max}$. Furthermore the upper component for $ {\kappa}
> 0$, ${\bf j}_{\tilde {\ell} + 1}$, does not have a zero in the $n = 0$
>branch but does have a zero in the
other branches so the radial quantum number is then $n_r = n - 1 =
0,1,\dots$. This means that the orbit with $n_r = 0,
{\kappa} < 0$ does not have a
pseudo - spin partner (in fact, this orbital is the "intruder" orbital
observed in heavy nuclei), but
the orbits with
$n_r, {\kappa}< 0, n_r -1, {\kappa} > 0$ are in the same branch and are
thus pseudo - spin partners
which agrees with experiment. Also we see from Fig. 2 that the RHS for $
{\kappa}> 0$ intersects the
LHS at a smaller Z than $ {\kappa}< 0$ and thus $E_{n_r - 1,{\kappa} > 0}
> E_{n_r,{\kappa} < 0}$ for the same
$\tilde{\ell}$ in agreement with experiment. Furthermore, as the RHS for
both $ {\kappa}$ increases,
they intersect the LHS at points in which the LHS has a larger slope and
therefore the points of
intersection, $Z_{n,{\kappa}}$, are closer. Thus these pseudo - spin
partners become closer in energy as the radial quantum number
increases.

These features can be seen in the limit of a large scalar potential, V$_s
\lsim 1$. In this limit, the RHS of (\ref {bessel>},
\ref {bessel<}) is large. We use the Bessel function identity \cite {wat}:

\begin{equation}
-\frac {{\bf j}_{\tilde{\ell} + 1} (Z) }{{\bf j}_{\tilde{\ell} }(Z)} =
\sum_{p=1}^{\infty}\
\big[\ \frac{1}{Z - Z_{p,\tilde{\ell}}^{(0)}} + \frac{1}{Z +
Z_{p,\tilde{\ell}}^{(0)}}\ \big],
\label {watson}
\end{equation}
which, in the $nth$ branch and for the LHS large and positive, can be
appoximated as
$-{ {{\bf j}_{\tilde{\ell} + 1} (Z) } / {{\bf j}_{\tilde{\ell} }(Z)}
\approx {1} /{ (Z - Z_{n,\tilde{\ell}}^{(0)}})}$. If, in addition,
$Y^{(0)}_{n,\tilde{\ell}}$ is large,
\begin{equation}
E_{n_r - 1,\kappa > 0} \approx E^{(0)}_{n,\tilde{\ell}}\ ,
\end{equation}
where we have denoted $E^{(0)}_{n,\tilde{\ell}}, Y^{(0)}_{n,\tilde{\ell}}$
as the values of $E,Y$ for
$Z = Z_{n,\tilde{\ell}}^{(0)}$, and
\begin{equation}
E_{n_r - 1,{\kappa} > 0} - E_{n_r,{\kappa} < 0}
\approx
\frac {(2{\tilde{\ell} } + 1)\ \Delta\ E_{n,\tilde{\ell}}^{(0)}
}{(Y^{(0)}_{n,\tilde{\ell}})\ ^2\
(\Delta - E_{n,\tilde{\ell}}^{(0)})} .
\label {diff}
\end{equation}
Thus we see that the energy splitting decreases as the binding energy
decreases, which is consistent
with the fact that pure pseudo - spin symmetry occurs when there are no
bound Dirac valence states, and
that the splitting decreases as the radial quantum number increases.
Furthermore, for states within the same
major shell, the splitting decreases as the pseudo - orbital angular
momentum decreases.

Hence, we have shown that pseudo-spin quasi-degeneracy in heavy nuclei can
be explained by the fact that nucleons in
a nucleus move in an attractive
scalar, -V$_s$, and repulsive vector, V$_v$, relativistic mean fields,
which are nearly
equal in magnitude,V$_s\sim\ $V$_v$. The energy splitting between states
with the same pseudo-orbital angular
momentum, $\tilde{\ell}$, decreases as the binding energy decreases and as
$\tilde{\ell}$ decreases.
Although such a near equality of
mean fields has been derived in specific relativistic field theories
\cite{wal,mad}, this result  probably is a
general feature of any relativistic model which fits nuclear binding
energies, and hence
very likely a general feature independent of any one model \cite {furn}. In
\cite {mad}, it was shown
that V$_s \sim $ V$_v$ for the isoscalar part of the nuclear
mean field (the largest part) but not for the isovector part, and the
isovector potential
has a different shape than the isoscalar potential. This implies that
pseudo - spin
symmetry may be enhanced in heavy proton - rich nuclei, with $N \sim Z$;
these nuclei
shall be measured in new radioactive beam facilitiies.

Pseudo - spin symmetry has been observed also in deformed nuclei\cite
{ben}; we are investigating the
deformed Dirac equation as well. Also, the explanation espoused in this
paper implies a connection
between the wavefunctions of the pseudo - spin doublets. This relationship
is being worked out.

The author thanks B. Serot for discussions, and N. Walet for making the
author aware of Ref \cite
{furn}.

\pagebreak

\pagebreak
\noindent\begin{figure}
\caption
{ The LHS of (\ref {bessel>}, \ref {bessel<}) plotted versus $Z$;
$Z^{(0)}_{n, \tilde
{\ell}}$ is the $nth$ zero of $j_{\tilde {\ell} }( Z^{(0)}_{n, \tilde
{\ell}}) = 0$.}

\end{figure}
\begin{figure}
\caption
{ The LHS of (\ref {bessel>}, \ref {bessel<}) (solid line), the RHS of
(\ref {bessel>}) (dashed
line), and the RHS of (\ref {bessel<}) (short dashed line) plotted versus
$Z$ for $\tilde
{\ell} = 1, V = 1.7,\Delta = 0.3, R= 33.5$, the radius of $^{208}Pb$ in
dimensionless units.}
\end{figure}

\end{document}